\documentclass{aastex}

\def\si{\sigma_i}
\def\ss{\sigma_s}
\def\Si{\Sigma_i}
\def\Ss{\Sigma_s}
\def\refnew#1{(\ref{#1})}

\usepackage{amssymb}
\usepackage{graphicx}

\begin{document}

\title{Spoke Formation Under Moving Plasma Clouds}
\author{Alison J. Farmer, Peter Goldreich}
\affil{Theoretical Astrophysics, MC 130-33, Caltech,
  Pasadena, CA 91125}
\affil{Institute for Advanced Study, Einstein Drive,
  Princeton, NJ 08540}
\email{ajf,pmg@ias.edu}
\altaffiltext{0}{14 pages, no tables, 2 figures\vspace{10in}\\
\emph{Running head:} Spoke formation under moving plasma clouds\\
\emph{Corresponding author:} Alison Farmer, MC 130-33, Caltech, Pasadena, CA
91125, ajf@ias.edu}

\begin{abstract}Goertz and Morfill (1983) propose that spokes on Saturn's
  rings form under radially moving plasma clouds produced by meteoroid
  impacts. We demonstrate that the speed at which a plasma cloud can
  move relative to the ring material is bounded from above by the
  difference between the Keplerian and corotation velocities. The
  radial orientation of new spokes requires radial speeds that are at
  least an order of magnitude faster. The model advanced by Goertz
  and Morfill fails this test.\end{abstract}


\newpage
\section{Introduction}

The nature of the ``spokes'' in Saturn's rings remains a matter of
speculation 25 years after their discovery by the Voyager
spacecraft. A brief summary of their properties is given here; for
more details see Mendis \emph{et al.} (1984) and references therein.
\begin{enumerate}
\item Spokes are transient radial albedo features superposed on
Saturn's rings.
\item Spokes are composed of dust with a narrow size distribution
  centered at 0.5 micron.
\item Spokes have optical depths of about 0.01.
 \item Spokes are only seen near corotation, which is where the
Keplerian angular velocity of the ring particles matches the planet's
rotational angular velocity. Corotation occurs in the outer B ring.
\item Individual spokes measure about 10,000 km in radial length and
2,000 km in azimuthal width; they extend over about 10\% of the ring
radius.
\item Spokes are seen preferentially on the morning ansa of Saturn's
  rings, and are most closely radial there. 
\item Spokes fade and are distorted by differential rotation
as they move from morning toward evening ansa.
\item A few observations have been interpreted as showing the birth of
individual spokes within 5 minutes along their entire lengths.  This
timescale implies a propagation velocity of at least 20 km s$^{-1}$.
\item Spokes are only observed at small ring opening angle to the Sun
  (McGhee \emph{et al.} 2005).

\end{enumerate}
The above observations suggest that spoke formation involves the
sudden lifting of a radial lane of dust grains from the surface of the
rings. Their subsequent fading and distortion is compatible with the
elevated dust grains moving on Keplerian orbits that intersect the
ring plane half an orbital period (i.e. about 5 hours) later.

Currently the most popular model for spoke formation is that of Goertz
and Morfill (1983, GM). GM propose that the formation of a spoke is
initiated when a meteoroid impacts the ring and creates a dense plasma
cloud. Electrons from the cloud are absorbed by the ring producing a
large electric field which levitates negatively charged dust
grains. The grains enter the cloud where they absorb additional
electrons. Overall charge neutrality is maintained by the net positive
charge of the plasma.

Because the dust grains are massive, they move on Keplerian
orbits. The plasma in which they are immersed is however tied to the
magnetic field lines which pass through the ionosphere of Saturn.  The
motion of the negatively charged dust relative to the positively
charged plasma produces an azimuthal electric field that causes the
plasma cloud to drift radially. GM argue that the plasma cloud will
continue to levitate dust grains as it moves.  According to their
calculations, the plasma cloud drifts in the radial direction, away
from corotation, at 20 -- 70 km/s. This velocity is sufficient to
account for the formation of a radial spoke of length 10,000 km within
5 minutes.

We perform a self-consistent calculation of the plasma cloud drift
velocity in \S \ref{sec:calc}. It establishes that the drift velocity
cannot exceed the difference between the local Keplerian and
corotation velocities. This upper limit is of order 1 km s$^{-1}$ in the
region where spokes are observed. We reveal the source of GM's error
in \S \ref{sec:GMprob} and estimate the correct drift velocity in \S
\ref{sec:recalc}. In \S \ref{sec:disc} we comment on the response of Morfill
\& Thomas (2005) to these issues. A short summary in \S \ref{sec:conc} concludes our
paper.

\section{Drift of a Plasma Cloud}
\label{sec:calc}

\begin{figure}
\includegraphics[width=1\textwidth]{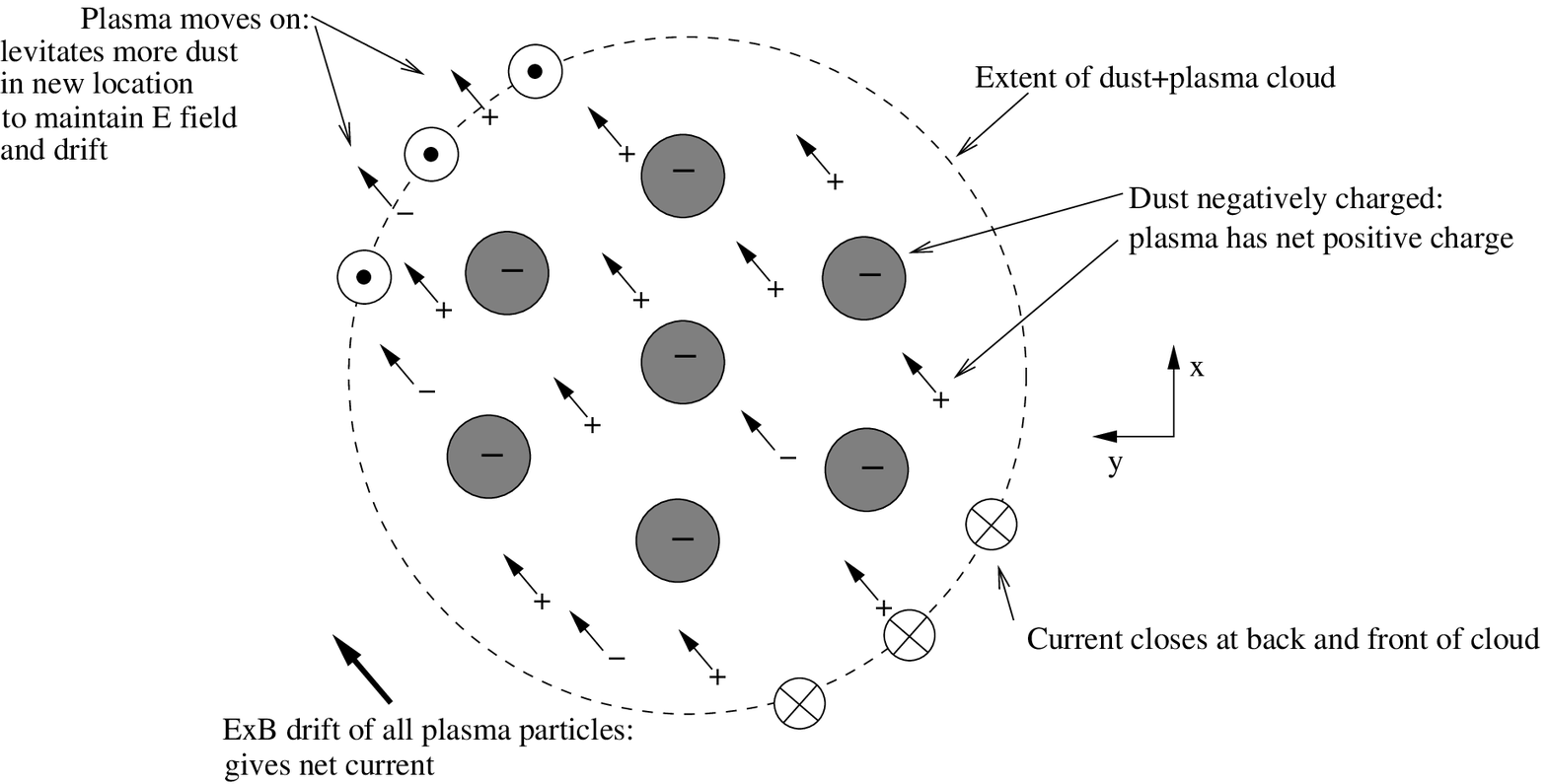}
\caption{The dusty plasma cloud, viewed from above in the local
rest frame of the ring particles.}
\label{fig:dust}
\end{figure}

The essence of GM's model is illustrated in Fig. \ref{fig:dust}. The
part of the ring plane at the base of the plasma cloud has a finite
Hall conductivity: in the presence of an electric field in the local
rest frame of the ring particles, the net positively charged plasma will
drift relative to the negatively charged dust grains which are embedded
within it. Away from the plasma cloud the ring plane has negligible
electrical conductivity.

Overall charge neutrality of the dusty plasma implies that currents
must close. Although the magnetospheric plasma maintains as
equipotentials the magnetic field lines linking the ring plane to the
ionosphere, it cannot carry currents across the field lines. Thus
currents that flow through the base of the plasma cloud must close in
Saturn's ionosphere. In the wake of the cloud, the levitated dust
rapidly combines with the positive ions that balance its charge,
leaving the spoke trail behind the cloud non-conducting.

\subsection{The model}
\label{sec:model}
\begin{figure}
\includegraphics[width=1\textwidth]{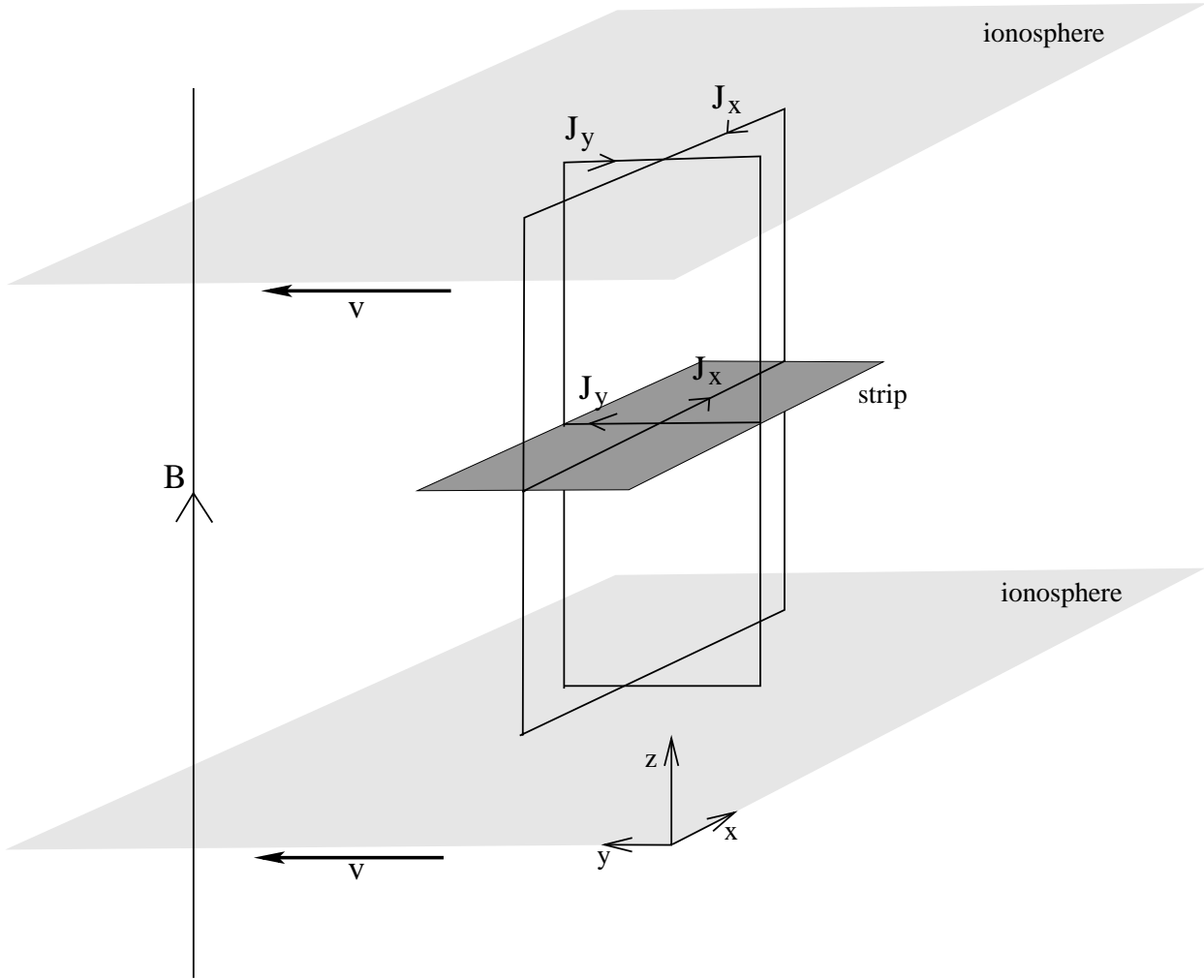}
\caption{The ionosphere-strip configuration described in \S \ref{sec:model}}
\label{fig:model}

\end{figure}

We analyse a simple model that captures the relevant features of the
system of a dusty plasma cloud, ionosphere, and magnetosphere. A 2D
strip of material in the $xy$-plane at $z=0$ represents the dust
grains (plus neutralizing ions) at the base of the plasma cloud at a
given time.\footnote{Defined in this way, the strip resembles a strip
of metal, in which the dust grains are analogous to the ion lattice
and the neutralizing ions are like the electrons which balance the
charge on the lattice. } Infinite sheets of a different material
at $z=\pm L$ take the place of the ionosphere. The magnetosphere
consists of a vertical magnetic field of strength $B$ embedded in a
massless plasma that maintains the field lines as equipotentials. All
calculations are done in the rest frame of the strip relative to
which the ionosphere moves with velocity ${\mathbf v}=(0,v,0)$ (see
Fig. \ref{fig:model}). Thus $x$ represents the radial direction in
the rings, and $y$ is azimuthal.

We wish to determine the drift velocity of the plasma in and above the
strip. This is equivalent to finding the horizontal electric field
in the system since the drift velocity ${\mathbf v_p}$ measured in the
same frame as $\bf{E}$ satisfies 
\begin{equation}
{\bf v}_p=c\frac{{\bf E}\times {\bf B}}{B^2}\, .
\label{eq:drift}
\end{equation}

\subsection{Finding the drift velocity}

In the rest frame of a conducting sheet, the height-integrated current
density ${\mathbf J}$ perpendicular to the magnetic field is given by
\begin{equation}
{\mathbf J}=\sigma {\mathbf E}+\Sigma \frac{{\mathbf B}\times {\mathbf E}}{|B|},
\label{eq:J}
\end{equation}
where $\sigma$ and $\Sigma$ are, respectively, the height-integrated
direct (Pedersen) and Hall conductivities. Because the ionosphere consists of
two sheets in parallel, its effective conductivity is twice that
of a single sheet.

Subscripts $s$ and $i$ are used to denote properties of the strip and
ionosphere. The electric field in the rest frame of the ionosphere is
related to that in the rest frame of the strip by
\begin{equation}
E_{xi}=E_{xs}+\frac{vB}{c}\, , \;\;  E_{yi}=E_{ys}.
\end{equation}
We assume, as GM implicitly did, that the currents are small enough so
as not to significantly perturb the externally imposed magnetic field
$B$, i.e. $4 \pi J/c \ll B$. We also take $v \ll c$.

Then, using Eq. \refnew{eq:J} for the currents in the rest frame of each
conductor, and expressing current conservation by 
\begin{equation}
{\mathbf J_s}+{\mathbf J_i}=0\, ,
\label{eq:curcon}
\end{equation}
 we solve the resulting simultaneous equations to give 
\begin{equation}E_{xs} = -\frac{vB}{c} \left[\frac{\Sigma_i \Sigma_t + \sigma_i
  \sigma_t}{\Sigma_t^2 + \sigma_t^2}\right],\end{equation}
\begin{equation}E_{ys} = \frac{v|B|}{c}\left[\frac{\Ss \si - \Si
  \sigma_s}{\Sigma_t^2 + \sigma_t^2}\right],\end{equation}
where $\Sigma_t = \Sigma_s+\Sigma_i$ and $\sigma_t =
\sigma_s+\sigma_i$.

The velocity ${\mathbf v}_p$ at which the plasma cloud moves then
follows from Eq. \refnew{eq:drift}: 
\begin{equation}v_{px} = v\left[\frac{\Ss \sigma_i - \Si
  \sigma_s}{\Sigma_t^2 + \sigma_t^2}\right]b,\label{eq:vpx}\end{equation}
\begin{equation}v_{py} = v \left[\frac{\Sigma_i \Sigma_t + \sigma_i
  \sigma_t}{\Sigma_t^2 + \sigma_t^2}\right]\,  ,
\label{eq:vpy}
\end{equation} 
where we have introduced $b=B/|B|=\textrm{sign}(B)$. 

The main message from Eqs. \refnew{eq:vpx} and \refnew{eq:vpy} is that
$|v_{px}|,|v_{py}| < v$.\footnote{Because Hall conductivities can be
negative, we can envisage the contrived case $\Ss\approx-\Si$, in
which we can in principle have $|v_{px}|>v$. However, this requires the
unlikely cancellation of the two unrelated Hall conductivities to a
high degree of accuracy. Moreover, $|v_{px}|>v$ requires $\Si>\si$,
which is not the case for Saturn's ionosphere (see section
\ref{sec:recalc}).}  This finding contradicts the result obtained by
GM (i.e. $|v_{px}| = 20-70 \textrm{~km~s}^{-1}$), because in the region
in which spokes appear, $v \lesssim 1\textrm{~km~s}^{-1}$. The
reasons for this discrepancy are detailed in \S \ref{sec:GMprob}.

Another interesting limit is one in which \emph{either or both} of the
direct and Hall ionospheric conductivities are much larger than both
components of the strip's conductivity. In this case, $|v_{px}|\ll v$
and $v_{py}\approx v$; i.e., the plasma cloud basically corotates with
the ionosphere.

\section{Where GM Erred}
\label{sec:GMprob}

GM neglected several terms when solving equations analogous to our
Eqs. \refnew{eq:drift}--\refnew{eq:curcon}. The components of the
height integrated currents in both the strip and ionosphere as a
function of the electric fields in their respective rest frames satisfy:
\begin{eqnarray}
J_{xs}&=&[\ss E_{xs}]-[b\Ss E_{ys}]\label{eq:serious}\\
J_{ys}&=&[\ss E_{ys}]+b\Ss E_{xs}\\
J_{xi}&=&\si E_{xi}-[b\Si E_{yi}]\\
J_{yi}&=&\si E_{yi}+[b\Si E_{xi}]\, .
\end{eqnarray}
Terms left out in GM's analysis are enclosed in square brackets. Since
the strip has negligible direct conductivity, neglecting terms
proportional to $\ss$ does no harm. The Hall current in the ionosphere
should be included since the Hall conductivity is comparable to the
direct conductivity in the ionosphere. However, this  neglect is not a
large source of error. The serious omission is that of the strip's
Hall current from $J_{xs}$, as indicated in Eq. \refnew{eq:serious}.

The above equations, with the bracketed terms omitted, yield the
incorrect result 
\begin{equation}
v_{px}=v \frac{\Ss}{\si}b \, ,
\end{equation}
from which $|v_{px}|>v$ follows for $|\Ss|>\si$.

Physically, GM's error is described by their incorrect statement that
``the motion of the plasma cloud does not constitute a perpendicular
(to ${\mathbf B}$) current as the plasma cloud is charge
neutral''. Because the dust grains are negatively charged the plasma
must have a net positive charge.  The azimuthal electric field set up
by the differential motion of the plasma and the dust grains does
cause the plasma to drift radially, but because the plasma is
net positive, this constitutes a radial Hall current. This radial
current closes in the ionosphere, and thus modifies the radial
electric field in both the ionosphere and the strip. But the radial
electric field is responsible for driving the azimuthal Hall current
in the strip, and this modification is not taken into account in GM.
They therefore incorrectly fix $v_{py}=v$, leading to a severe overestimate
of $v_{px}$.  All these factors are accounted for in the
self-consistent calculation outlined in \S \ref{sec:calc}.

\section{Recalculation of Drift Velocity}
\label{sec:recalc}

We recalculate the drift velocity of a plasma cloud from Eqs.
\refnew{eq:vpx} and \refnew{eq:vpy}.

The direct conductivity in the strip is very small because the dust
grains have low mobility and the plasma is tied to the magnetic field
lines. The Hall conductivity, which results from the ${\bf E\times B}$
drift of the positively charged plasma, is correspondingly high, of
order
\begin{equation}\Ss \sim -\frac{N_p e c}{|B|}\,  ,
\end{equation} 
where $N_p e$ is the height integrated charge density of the
plasma.\footnote{The Hall conductivity is negative because it is
defined as positive for the commonly encountered case in which
electrons are the dominant current carriers.} Using the approach of
GM, we estimate the height
integrated charge density to be of order $10^1$ esu cm$^{-2}$, which
gives $\Ss \sim -10^{14} \textrm{~cm~s}^{-1}$.\footnote{This number
  may be spuriously high, because the approach of GM gives more charge
  on the dust grains than was originally present in the plasma.}

Saturn's ionospheric conductivities vary with latitude and with time,
with typical dayside values of the height-integrated direct conductivity being $\si
\simeq 10^{12}-10^{13} \textrm{~cm~s}^{-1}$, and nightside values
about 100 times smaller (Cheng and Waite 1988). The Hall conductivity
is about an order of magnitude smaller, e.g. for the auroral region we
have $\si = 5 \times 10^{13}\textrm{~cm~s}^{-1}$ and $\Si = 8 \times
10^{12} \textrm{~cm~s}^{-1}$ (Atreya \emph{et al.} 1983). The direct current
is carried predominantly by protons, and the Hall current by
electrons, which suffer fewer collisons per gyroperiod than the
protons, so ${\bf E \times B}$-drift more freely. Collision frequencies of
electrons and protons in Saturn's ionosphere are smaller than the
respective gyrofrequencies, so the Hall conductivity is smaller than
the direct conductivity.

We substitute into Eqs. \refnew{eq:vpx} and \refnew{eq:vpy} the typical
dayside values $2 \si = 1 \times 10^{13}\textrm{~cm~s}^{-1}$ and
$2 \Si = 1 \times 10^{12}\textrm{~cm~s}^{-1}$ (the factors of two
account for both hemispheres of the ionosphere, although we note that
we do not expect north-south symmetry of conductivities). We then
obtain
\begin{equation}
v_{px}=6\times 10^{-2} v,\;v_{py}=-2 \times 10^{-3} v,
\end{equation}
where we have used $b=-1$ as is appropriate for Saturn, and where $v
\lesssim 1$ km/s. With these velocities, spokes will not form quickly
or radially. The strip Hall conductivity is high enough to drag the
plasma column in an almost Keplerian orbit.

\section{Discussion}
\label{sec:disc}
In response to this paper, Morfill \& Thomas (2005) revisit the GM
model for spoke formation. They provide further details about dust
charging and plasma cloud structure. These details are however
irrelevant to the criticisms raised here, and make no difference to
the fact that, just as the radial electric field produces an azimuthal
Hall current, an azimuthal electric field will produce a radial Hall
current. These currents and electric fields must be treated
self-consistently, resulting in the conclusions reached in this paper.

\section{Conclusions}
\label{sec:conc}

We have studied the physical situation in which a strip of conducting
material moves between two parallel sheets of a different material, to
which it is joined by perpendicular magnetic field lines (Fig.
\ref{fig:model}). The plasma on these field lines will drift in the
plane of the strip, both
parallel and perpendicular to the relative velocity vector. We have shown that the magnitude of this drift
velocity cannot exceed that of the relative strip-sheet velocity.

Application of this limit to the most popular model for spoke
formation demonstrates that the model rests upon a gross overestimate of the
velocity at which a plasma cloud can drift.

\section{Acknowledgments}

This research was supported by NSF grant AST 00-98301. We thank J. H. Waite for
information on Saturn's ionospheric conductivity.

\newpage

\newpage

\end{document}